\documentclass[aps,pre,onecolumn,showpacs,floatfix]{revtex4-1}
\usepackage{amssymb,amsmath,epsfig,graphicx}

\begin{document}

\title{\bf Open Problems on Information and Feedback Controlled Systems}

\author{F. J. Cao}
\email{francao@fis.ucm.es}
\affiliation{Departamento de F\'{\i}sica At\'omica, Molecular y Nuclear,
Universidad Complutense de Madrid, \\
Avenida Complutense s/n, 28040 Madrid, Spain}

\author{M. Feito}
\email{feito@fis.ucm.es}
\affiliation{Departamento de F\'{\i}sica At\'omica, Molecular y Nuclear,
Universidad Complutense de Madrid, \\
Avenida Complutense s/n, 28040 Madrid, Spain.}

\begin{abstract}     
Feedback or closed-loop control allows dynamical systems
to increase their performance up to a limit imposed by the second
law of thermodynamics. It is expected that within this limit, the
system performance increases as the controller uses more information
about the system. However, despite the relevant progresses made
recently, a general and complete formal development to justify this
statement using information theory is still lacking. We present here
the state-of-the-art and the main open problems that include aspects
of the redundancy of correlated operations of feedback control and
the continuous operation of feedback control. Complete answers to
these questions are required to firmly establish the thermodynamics
of feedback controlled systems. Other relevant open questions
concern the implications of the theoretical results for the
limitations in the performance of feedback controlled flashing
ratchets, and for the operation and performance of nanotechnology
devices and biological systems.
\end{abstract}

%\keyword{closed-loop control; information; thermodynamics}

\pacs{89.70.Cf, 05.20.-y, 05.40.-a, 02.30.Yy}

\maketitle

%%%%%%%%%%%%%%%%%%%%%%%%%%%%%%%%%%%%%%%%%%%%%%%
\section{Introduction}
%%%%%%%%%%%%%%%%%%%%%%%%%%%%%%%%%%%%%%%%%%%%%%%

Control over physical systems exerted by an external agent is
ubiquitous in physics and engineering, where it is used for purposes
such as stabilizing unstable dynamics or increasing system
performance. Studying the properties of controlled systems is a
fundamental task for engineering purposes and from a fundamental
physical point of view~\cite{bec05}.
\par

The operation of the controller modifies the system dynamics, and
the system in turn can also influence the controller decisions.
Controllers are called closed-loop or feedback controllers if their
actions are dependent on the state of the system and  called
open-loop if otherwise. In other words, closed-loop or feedback
controllers actuate the system using the information about the
system state, while open-loop controllers actuate the system
independently of the system state.

It is intuitively clear that using information about the system
state is potentially useful to improve the controller action on the
system and increase the system performance (e.g., the power output
or the efficiency). However, the ceiling of such performance
enhancement from making use of system state information is not
clearly quantified. In particular, a complete and general
development is lacking for calculating the performance in feedback
controlled systems using thermodynamic principles. The aim of this
paper is to present and put in context several open questions in
this field.

In the following section we briefly review some selected results
about the relation between information and feedback. Next, in
Section~\ref{sec:open}, we present several of the still unsolved
questions and discuss their implications. Finally, our general
conclusions are presented in Section~\ref{sec:conclusions}.

%%%%%%%%%%%%%%%%%%%%%%%%%%%%%%%%%%%%%%%%%%%%%%%
\section{State-of-the-Art} \label{sec:state}
%%%%%%%%%%%%%%%%%%%%%%%%%%%%%%%%%%%%%%%%%%%%%%%

In this section, we introduce some helpful results in the context of
the below-mentioned open problems. The results presented here will
be useful for one to get insight into the open questions and address
their answers.

In Subsection~\ref{sec:Maxwell} we comment some of the main results
found in the framework of Maxwell's demon and computation
theory~\cite{lef03, szi29, lan61, ben82, zur89, zur89b,ben73}. In
Subsection~\ref{sec:information} we summarize recent results about
the relation between the achievable decrease of entropy and the
information about the system gathered by the
controller~\cite{llo89,tou00,tou04}. In
Subsection~\ref{sec:feedback} we show recent results concerning
performance and information in a relevant stochastic system, namely,
a feedback flashing ratchet. In Subsection~\ref{sec:thermodynamics}
we summarize recent results on the thermodynamics of feedback
controlled systems. Finally, in Subsection~\ref{sec:generalization}
we comment several very recent results that generalize statistical
physics identities and relations to feedback controlled systems.

%%%%%%%%%%%%%%%%%%%%%%%%%%%%%%%%%%%%%%%%%%%%%%%
\subsection{Maxwell's Demon}\label{sec:Maxwell}
%%%%%%%%%%%%%%%%%%%%%%%%%%%%%%%%%%%%%%%%%%%%%%%

The trade-off between information and feedback control has been a
difficult task since the birth of Maxwell's demon~\cite{lef03}, a
hypothetical being that gathers information about a system to
decrease the entropy of it without performing work. Szilard's
seminal work~\cite{szi29} contains the basic ingredients of the
interrelation between information theory and thermodynamics, which
was later precisely stated in the Landauer's principle: The erasure
of one bit of information at temperature $T$ implies an energetic
cost of at least $k_BT\ln 2$~\cite{lan61}. Bennett~\cite{ben82}
pointed out that the Landauer's principle is essential to preserve
the second law of thermodynamics in feedback systems, as the
controller must erase its memory after each cycle to allow the whole
system to truly operate cyclically. How to achieve the shorter
description for the memory record of the demon in order to minimize
the energetic erasure cost was established by
Zurek~\cite{zur89,zur89b} by using an algorithmic complexity
approach. He proposed to compress the memory record of the demon
before the erasure. This compression of the memory can be done, in
principle, without energy expenditure---{\em i.e.}, reversibly. The
reversibility of the compression process is related to the fact that
a memory can be erased at zero energy cost provided its state is
known, {\em i.e.}, stored in another memory \cite{ben73}. This
implies that the minimum energy required to erase the memory of the
demon is not given by the number of bits stored, but by the minimum
number of bits in which the information contained in the memory can
be stored (after compression). This important result raises the
attention on the relevance of considering whether the information is
redundant or not.

%%%%%%%%%%%%%%%%%%%%%%%%%%%%%%%%%%%%%%%%%%%%%%%
\subsection{Information-Theoretic Limits of Control}\label{sec:information}
%%%%%%%%%%%%%%%%%%%%%%%%%%%%%%%%%%%%%%%%%%%%%%%

A fundamental magnitude in information theory~\cite{cov91} is the
Shannon entropy of a random variable $X$. If $X$ takes values $x$
with probability $p_X(x)$, then the Shannon entropy is given, in
nats ($ 1 \mbox{ bit} = \ln 2 \mbox{ nats}$), by
\begin{equation}
  H(X)=-\sum_{x}p_X(x)\ln p_X(x)
\end{equation}
and it is a measure of the average uncertainty of the random
variable~\cite{cov91}. (When $X$ is the macrostate of a system and
$x$ is its microstate, $H(X)$ is related with the statistical
physics entropy $S$ of the system by $S=k_BH$). On the other hand,
the conditional entropy of $X$ given another random variable $C$ is
\begin{equation}\label{hxc}
H(X|C)=\sum_{c}p_{C}(c)H(X|C=c)
\end{equation}
with $H(X|C=c)=-\sum_{x} p_{X|C}(x|c) \ln p_{X|C}(x|c).$ From these
magnitudes, the mutual information between two random variables,
which is the information that one variable contains about the other
one, is computed as
\begin{equation}
  I(X;C)=H(X)-H(X|C)
\end{equation}
Hence, the mutual information can be viewed as the reduction of
uncertainty of one of the random variables due to the knowledge of
the other~\cite{cov91}.
\par

Fundamental limits on the controllability of dynamical systems can
be studied in the light of information
theory~\cite{llo89,tou00,tou04}. Recently, it has been established
that the maximum amount of entropy $\Delta
H_{\mbox{\footnotesize{closed}}}$ that can be extracted from any
dynamical system by a single closed-loop actuation is upper bounded
by the maximum entropy decrease achievable by open-loop control
$\Delta H_{\mbox{\footnotesize{open}}}$ plus the mutual information
between the controller $C$ and the state of the system
$X$~\cite{tou00,tou04}:
\begin{equation}
\Delta H_{\mbox{\footnotesize{closed}}}\leq \Delta
H_{\mbox{\footnotesize{{open}}}} + I(X;C)
\end{equation}
This result has provided insight into the relations between
information and feedback, explicitly quantifying and showing one of
the limitations of attainable performance improvement with using
certain amount of system information. The analogous results for the
physical entropy of a physical system has been recently obtained for
repeated closed-loop actuations, see
Section~\ref{sec:thermodynamics} and~\cite{cao09}. Also recently the
application of information theory to biochemical reaction networks
has allowed to show the limits of the ability to control
fluctuations in molecular abundances inside the cells as a function
of the number of signaling events~\cite{les10}.

%%%%%%%%%%%%%%%%%%%%%%%%%%%%%%%%%%%%%%%%%%%%%%%
\subsection{Feedback Controlled Ratchets and Information}\label{sec:feedback}
%%%%%%%%%%%%%%%%%%%%%%%%%%%%%%%%%%%%%%%%%%%%%%%

\begin{figure}
\begin{center}
\includegraphics [scale=0.6] {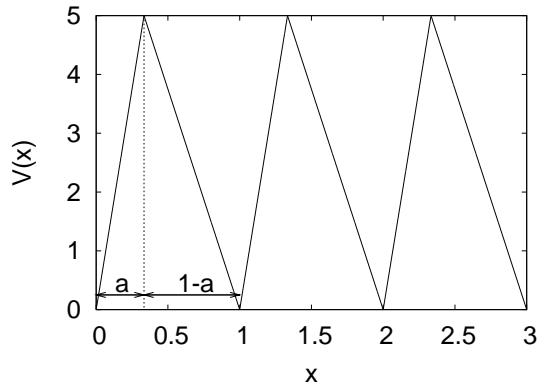}
\end{center}
\caption{In a feedback flashing ratchets a controller with
information of the  state of the system rectifies the thermal
fluctuations of Brownian particles  by switching on and off a
periodic potential (which can be asymmetric, as the one depicted in
the plot, or symmetric). This rectification of thermal fluctuations
generates particle flux even against an external force, thus
performing
work~\cite{cao04,cao04-1,cao04-2,cao04-3,cao04-4,cao04-5,cao04-6,cra08,cra08-1,lop08,ast98}.}
\label{fig:figure1}
\end{figure}

Let us now briefly review some related results applied to Brownian
ratchets, \emph{viz.}, stochastic systems that are capable of
rectifying thermal fluctuations by working out of equilibrium.
Recently, feedback controlled versions of Brownian ratchets have
been
studied~\cite{cao04,cao04-1,cao04-2,cao04-3,cao04-4,cao04-5,cao04-6,cra08,cra08-1,feedback,feedback-1,inertial,inertial-1,inertial-2,inertial-3,inertial-4}
and experimentally realized~\cite{lop08}. (See
Figure~\ref{fig:figure1} for a Brownian ratchet that is based on the
flashing of a potential.)

These feedback controlled ratchets can improve the performance over
its open-loop counterparts thanks to the information the controller
gets about the system state.\ These information-dependent ratchets
are relevant in chemical and biological
systems~\cite{zho96,bie07,ser07}, and also in nanotechnology
devices~\cite{cao04,cao04-1,cao04-2,cao04-3,cao04-4,cao04-5,cao04-6,cra08,cra08-1,lop08}.
Other ratchets and Brownian motors inspired by Maxwell's demon
present a directed motion when there is a temperature difference
between two different parts of the motor (one part could be thought
as the controller and the other as the controlled
system)~\cite{fey63,bro05}. However, for these later systems no
systematic study on the role of information has been performed. On
the other hand, for feedback flashing ratchets, relations between
the information $I$ (about the system) gathered by the controller
per step and the performance have been recently
obtained~\cite{cao07,fei07}. These relations have been established
for a collective feedback flashing ratchet compounded of one or few
particles, considering the optimal protocol for one particle and its
generalization to few particles. In this protocol the controller
gathers the sign of the net force over the particles if the
potential would be on. If the sign is measured with a perfect
measurement device and it is transmitted to the controller through a
noiseless channel, the controller gets one bit of information about
the system (the maximum value of the information $I$ per step for
this system). However, if the measurement or the transmission is
imperfect, the controller gets less than one bit of information
about the system, and this limits the increase of performance
improvement that can be obtained thanks to the information.

Let us call $v_{\mbox{\footnotesize{open}}}$ the maximum flux that
can be achieved with an open-loop control protocol, and
$v_{\mbox{\footnotesize{closed}}}$ the maximum flux that can be
obtained with a closed-loop control protocol that uses an amount of
information $I$ per measurement step. The computations
in~\cite{cao07} have shown that the maximum increase in the flux
that can be obtained has an upper bound proportional to the square
root of the information $ I $ received in each step, \emph{i.e.},
\begin{equation} \label{vbound}
    v_{\mbox{\footnotesize{closed}}}-v_{\mbox{\footnotesize{open}}}\leq
    C_1 \sqrt{I}
\end{equation}
(with $C_1$ being a constant depending on the system's
characteristics).

An analogous bound has been found for the maximum power output
in~\cite{fei07}. Let us call $ P_{\mbox{\footnotesize{open}}} $ the
maximum power output that can be obtained with an open-loop control,
and $ P_{\mbox{\footnotesize{closed}}} $ the maximum power output
that can be obtained with a closed-loop control that uses an amount
of information $ I $ per measurement step. It has been shown
in~\cite{fei07} that the maximum increase in the power output has an
upper bound proportional to the information $I$ gathered by the
controller per step, \emph{i.e.},
\begin{equation} \label{Pbound}
  P_{\mbox{\footnotesize{closed}}}-P_{\mbox{\footnotesize{open}}}\leq C_2 I
\end{equation}
(with $C_2$ being a constant depending on the system's
characteristics).

The previous bounds [Equations~(\ref{vbound}) and (\ref{Pbound})]
provide insight into the limitations of performance enhancement that
can be obtained with a given amount of information.\ In particular,
they have already been used to assess the feasibility of an
experimental realization of feedback ratchets before its
implementation \cite{cra08,cra08-1,lop08}. In addition, these bounds
are an example of the type of predictions that a general theory
would provide about the relations between information and feedback
control.

%%%%%%%%%%%%%%%%%%%%%%%%%%%%%%%%%%%%%%%%%%%%%%%
\subsection{Thermodynamics of Feedback Controlled Systems}\label{sec:thermodynamics}
%%%%%%%%%%%%%%%%%%%%%%%%%%%%%%%%%%%%%%%%%%%%%%%

Correlated repeated actuation of the feedback controller is present
when the controller repeatedly actuates the system without waiting
for the relaxation of the system to equilibrium between actuations.
(Thus, adiabatic approximations~\cite{all08} cannot be applied.)\ In
the correlated repeated actuation, the information received by the
controller from the different measurements has redundancies.\ These
redundancies have to be consistently taken into account as they
reduce the possibilities to increase the performance up to the
bounds that would be predicted from the known results for
uncorrelated repeated actuations~\cite{lef03,szi29,lan61,ben82}.

Recently, the entropy reduction of a system due to the correlated
repeated actuation of an external controller has been
computed~\cite{cao09}. This was the main lacking ingredient to
establish the thermodynamics of feedback controlled systems
(\emph{i.e.}, to be able to compute the thermodynamic potentials and
its relations for these systems). In the following paragraphs we
include an excerpt of~\cite{cao09} summarizing the main~results.

A control step of a controller involves: measuring the system,
deciding the control action to take given the measurement output,
and acting on the system following the control action chosen. From
the point of view adopted in~\cite{cao09}, \emph{i.e.}, the point of
view of the system, we are just concerned about which action the
controller has taken on the system, whose consequence is to modify
the evolution of the system. We are not concerned about the state of
the controller.

Let us call $X_k:=X(t_k)$ the macrostate of a general physical
system at the $k$th control step of the controller (at time $t_k$),
and $C_k$ the action taken by the controller (it is \emph{not} a
specification of the state of the controller). The entropy just
before the first measurement is given by the statistical entropy
\begin{equation}
  S^b_1=-k_B\sum_{x\in{\cal X} }p_{X_1}(x) \ln p_{X_1}(x) =: k_B H(X_1)
\end{equation}
with ${\cal X}$ the set of possible microstates of the system. If
the measurement implies the control action $C_1=c$, the entropy of
the system decreases to
\begin{equation}
k_B H(X_1|C_1=c):=- k_B \sum_{x\in{\cal X}} p_{X_1|C_1}(x|c) \ln
p_{X_1|C_1}(x|c)
\end{equation}
because knowing that the control action is $C_1=c$ gives additional
information about the actual microstate of the system, due to the
correlation between measurements by the controller and control
actions. Therefore, the average entropy \emph{after} the first
control step can be obtained by the following weighted average over
the set ${\cal C}$ of all possible control actions:
\begin{equation}
  S^a_1=\sum_{c\in{\cal C} }p_{C_1}(c) k_B H(X_1|C_1=c) =: k_B H(X_1|C_1)
\end{equation}
Hence the average variation of the entropy at the first step is
\begin{equation} \label{SinfoOne}
    \Delta S_1 = S^a_1-S^b_1= k_B H(X_1|C_1) - k_B H(X_1)=: - k_B I(X_1;C_1)
\end{equation}
\emph{i.e.}, it is $k_B$ times the (minus) mutual
information~\cite{cov91} between $X_1$ and $C_1$ given in nats (ln 2
nats = 1~bit).

After the first measurement we have a set of macrostates (one for
each possible control action). Each of the macrostates evolves to
give an entropy $k_B H(X_2|C_1=c)$ just before the second control
step. The average entropy of the system after the second step is
\begin{equation}
  S^a_2  = \sum_{c,c^\prime\in{\cal C}} p_{C_2C_1}(c^\prime,c)
 k_B H(X_2|C_2=c^\prime,C_1=c) = k_B H(X_2|C_2,C_1)
\end{equation}
and the average variation of the entropy at this second control step
is
\begin{equation}
\Delta S_2 =S^a_2-S^b_2=k_B H(X_2|C_2,C_1)-k_B H(X_2|C_1)=-k_B
I(X_2;C_2|C_1)
\end{equation}
This conditioning of the mutual information shows that the entropy
of the system is only reduced by new non-redundant information.
Analogously we get for the average entropy reduction of the $k$-th
step
\begin{equation}\label{sk2}
\Delta S_k = -k_B I(X_k;C_k|C_{k-1},\ldots,C_1)= - k_B
H(C_k|C_{k-1},\ldots,C_1)
 + k_B H(C_k|C_{k-1},\ldots,C_1,X_k)
\end{equation}
Therefore, the \emph{total average entropy reduction due to the
  information used} in $M$ control steps is \linebreak $\Delta S_{\rm
  info}=\sum_{k=1}^{M}\Delta S_k$, \emph{i.e.},
\begin{equation}\label{Sinfo}
  \Delta S_{\rm info}=-k_B\sum_{k=1}^{M}I(C_k;X_k|C_{k-1},\ldots,C_1)
\end{equation}
or equivalently
\begin{equation}\label{Sinfob}
  \Delta S_{\rm info}=-k_B H(C_M,\ldots,C_1)+k_B\sum_{k=1}^{M} H(C_k|C_{k-1},\ldots,C_1,X_k)
\end{equation}
The first term in this last equation gives the maximum entropy
decrease attainable, and the second term gives the reduction on this
maximum due to decorrelations between the control actions and the
state of the system (decorrelations that can be due to errors and
noises affecting the controller operation). These results have a
broad range of applications  including deterministic and
non-deterministic feedback controls, and they indicate that the
average entropy reduction due to the information used can be
computed in terms of the joint probabilities for the state of the
system and the control actions history.

\begin{figure}
\begin{center}
\includegraphics [scale=0.6] {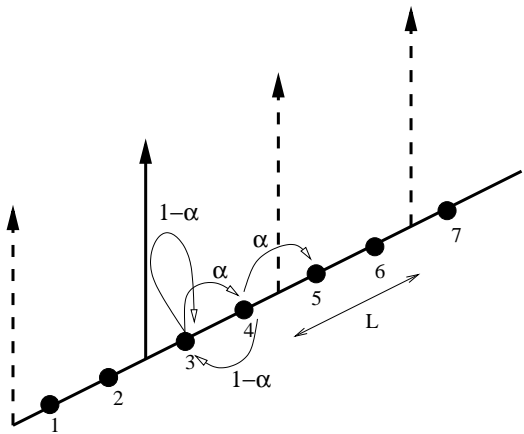}
  \end{center}
\caption{Markovian particle pump with two lattice sites between
    barriers. The particle is in contact with a thermal bath at temperature
    $T$, and an external controller measures the particle location and raises
    the closest barrier to the left. The result is that the particle is
    further and further constrained to the right, increasing its potential
    energy. This is a simple feedback controlled system that extracts useful
    work from the entropy reduction due to the information about the
    system used by the external feedback controller~\cite{cao09}.}
\label{fig:figure2}
\end{figure}

These results have also allowed us to establish the thermodynamics
of feedback controlled systems. As an application, we have computed
the efficiency of isothermal feedback controlled system, and as an
example, we have applied the results to compute the efficiency of a
Markovian particle pump. This isothermal feedback controlled system
can extract work from a single heat bath, thanks to the use of
information, see Figure~\ref{fig:figure2} and~\cite{cao09}. For an
experimental demonstration of a Markovian particle pump,
see~\cite{toy10}. Very recently other motors that extract work from
a single heat bath have been
studied~\cite{abr11,abr11-1,abr11-2,abr12}.

%%%%%%%%%%%%%%%%%%%%%%%%%%%%%%%%%%%%%%%%%%%%%%%
\subsection{Generalization of Identities and Relations to Feedback
  Controlled Systems}
\label{sec:generalization}
%%%%%%%%%%%%%%%%%%%%%%%%%%%%%%%%%%%%%%%%%%%%%%%

Very recently, several important identities and nonequilibrium
equalities of Statistical Physics have been generalized to enlarge
their range of applicability to feedback systems. Sagawa and
Ueda~\cite{sag12,sag12-1,sag12-2,sag12-3} have extended the
fluctuation theorem and the Jarzynski relation~\cite{jar97} to
feedback controlled systems even for multiple measurements. Once
again, they show that the mutual information plays a crucial role to
reformulate the nonequilibrium statistical mechanics to feedback
controlled processes. Abreu and Seifert have extended the Hatano and
Sasa equality to feedback controlled processes~\cite{abr12},
considering systems that, even for fixed control parameters, have
nonequilibrium steady states with dissipation. These results also
provide bounds for the power output that can be extracted under
isothermal conditions. The detailed fluctuation theorem has also
been generalized to processes with feedback
in~\cite{hor10,pon10,lah12}. All these results broadly extend the
available resources to study feedback controlled systems and to find
the bounds to their performance.

%%%%%%%%%%%%%%%%%%%%%%%%%%%%%%%%%%%%%%%%%%%%%%%
\section{Open Problems} \label{sec:open}
%%%%%%%%%%%%%%%%%%%%%%%%%%%%%%%%%%%%%%%%%%%%%%%

In this section we present several open problems in the context of
the connections between the study of feedback controlled systems and
information theory. We have selected those questions that we think
would imply a greater progress in the understanding of these
connections, and would further advance the understanding of feedback
controlled systems. The questions we have selected mainly concern
the establishment of bounds to the performance and the establishment
of the thermodynamics of feedback controlled systems. Other open
questions on the links between feedback control and information
theory are raised in the Bechhoefer review on control~\cite{bec05}.
Regarding other open questions that concern the implications of the
theoretical results for the performance limitations of particular
systems, due to their particular relevance we discuss here feedback
controlled flashing ratchets, and also, nanotechnology devices and
biological systems.

%%%%%%%%%%%%%%%%%%%%%%%%%%%%%%%%%%%%%%%%%%%%%%%
\subsection{Feedback Controlled Systems: Limitations to the Performance and
Thermodynamics}
%%%%%%%%%%%%%%%%%%%%%%%%%%%%%%%%%%%%%%%%%%%%%%%

Feedback controlled systems are usually designed with the aim that
the controller profits from the information about the system state
it receives in order to increase the performance of the system.
Therefore, the first natural main question is how much the
performance can be increased with a certain amount of information,
or, stated otherwise, which are the bounds of the performance
increase. The second main question is the establishment of the
thermodynamics of feedback controlled systems accounting for
thermodynamic effects of the information transfer. This second
question can be seen as included in the first one, as, for instance,
the computation of the maximum efficiency involves a bound in the
work that can be extracted from the feedback controlled system.

The case of a single operation of a feedback controller has been
extensively studied in the context of Maxwell's demon and in
computation theory. In this case the main thermodynamical results
are known; see Sections~\ref{sec:Maxwell} and \ref{sec:information}.

On the other hand, the study of the repeated operations of a
feedback controller presents new ingredients. We have to distinguish
whether the operations of the feedback controller are correlated or
uncorrelated. In the latter case, the system has lost memory of the
previous actuation of the controller when its status is measured
again to decide the next actuation by the controller. Therefore, the
uncorrelated operations of the feedback can be decomposed as a
succession of independent single operations, and thus the known
previous results can be applied. However, the case of correlated
repeated operation of the feedback controller has fundamental new
ingredients that must be taken into account.

%%%%%%%%%%%%%%%%%%%%%%%%%%%%%%%%%%%%%%%%%%%%%%%
\subsubsection{Correlated Repeated Actuation of the Feedback
  Controller} \label{sec:correlated}
%%%%%%%%%%%%%%%%%%%%%%%%%%%%%%%%%%%%%%%%%%%%%%%

In~\cite{cao09} (see Section~\ref{sec:thermodynamics}) the entropy
reduction due to the repeated operation of the controller has been
derived from the point of view of the feedback controlled system.
This point of view is also used to derive the result
of~\cite{llo89,tou00,tou04} (see Section~\ref{sec:information}).
From this point of view, the controller is an external agent and we
are not concerned about its internal state. We are only concerned
about the effects on the system due to its interaction with the
controller. One of the effects is the reduction of the system's
entropy due to the additional determination of the system's
macrostate through the information obtained from the system by the
controller.

Therefore, there are still open questions to be answered. One
particular open question involves the precise relations between
these recent results and the algorithmic complexity approach adopted
by Zurek~\cite{zur89,zur89b} to study the correlated repeated
actuation of a feedback controller from the perspective of the
system plus controller. This study has allowed Zurek to establish
the minimum energetic cost for erasing a memory that has stored the
measurements the controller has processed (see
Section~\ref{sec:Maxwell}). This minimum energetic cost also sets
thermodynamic constraints that lead to a maximum entropy reduction
for the system.

Another open question is the application or the extension of the
results found in~\cite{cao09} to find bounds for other performance
measures of feedback controlled systems, for example, the maximum
attainable flux or power.

%%%%%%%%%%%%%%%%%%%%%%%%%%%%%%%%%%%%%%%%%%%%%%%
\subsubsection{Continuous Actuation of the Feedback Controller}
\label{sec:continuous}
%%%%%%%%%%%%%%%%%%%%%%%%%%%%%%%%%%%%%%%%%%%%%%%

Continuous actuation of the feedback controller is effectively
present when the controller repeatedly actuates on the system with a
period shorter than any of the characteristic time scales of the
system dynamics. It is therefore a limiting case of the correlated
repeated actuation discussed above. However, the continuous time
limit of the results found in~\cite{cao09} is not direct because of
the nontrivial relationship between differential entropy and
discrete entropy (see~\cite{cov91}). Therefore, even the extension
of the results in~\cite{cao09} to the continuous actuation limit
remains still an open problem.

Some techniques of filtering theory might be useful for the study of
the information and entropy flows associated with continuous-time
processes in non-equilibrium statistical mechanical
systems~\cite{mit05}. In addition, methods of optimal control theory
have been proven useful to compute limits on how much work can be
extracted from linear systems~\cite{san07}.

%%%%%%%%%%%%%%%%%%%%%%%%%%%%%%%%%%%%%%%%%%%%%%%
\subsection{Feedback Controlled Flashing Ratchets: Limitations to the
Performance and Thermodynamics} \label{sec:feedbackopen}
%%%%%%%%%%%%%%%%%%%%%%%%%%%%%%%%%%%%%%%%%%%%%%%

Feedback flashing ratchets are relevant feedback controlled systems
that would provide relevant applications of the general results
obtained for feedback controlled systems. In addition, they can
provide enlightening examples to guide the progress in the solution
of the previous open problems.

The application of the results of~\cite{cao09} to feedback
controlled flashing ratchets has not been done yet. It will allow to
establish bounds on the entropy reduction attainable for feedback
flashing ratchet and to establish the thermodynamics of these
systems, which are theoretically and technologically relevant.
Therefore, this is another relevant open problem.

Previously, in Section~\ref{sec:feedback}, we have summarized the
bounds that have been found in~\cite{cao07,fei07} for the
performance increase thanks to the information used by the feedback
controller, when the performance is measured by the
flux~\cite{cao07} or by the power output~\cite{fei07}. These
relation were stated in terms of the information per actuation.
However, how to restate these relations in terms of the total amount
of information per unit time ({\em i.e.}, only computing
non-redundant information per unit time) is still an open question.
This restatement will help to put these results~\cite{cao07,fei07}
in relation with those found in~\cite{cao09} for the entropy
reduction and the thermodynamics of a general feedback controlled
system.

%%%%%%%%%%%%%%%%%%%%%%%%%%%%%%%%%%%%%%%%%%%%%%%
\subsection{Limitations to the Operation of Nanotechnology Devices and
  Biological Systems}
\label{sec:limitations}
%%%%%%%%%%%%%%%%%%%%%%%%%%%%%%%%%%%%%%%%%%%%%%%

In many nanotechnology devices and biological systems, one is
interested in increasing the performance of one part of the system
through the bidirectional interaction of other part of the
system~\cite{cao04,cao04-1,cao04-2,cao04-3,cao04-4,cao04-5,cao04-6,cra08,cra08-1,lop08,zho96,bie07,ser07}.
Therefore, we can think of the subsystem whose performance we want
to increase as a feedback controlled system, and consider the other
interacting part as the controller. Thus, the bounds of the
performance of the feedback controlled systems will imply
limitations to the operation of nanotechnology devices and
biological systems. Finding these implications is still a very
relevant open question.

An illustrative example of this type of questions is the previously
mentioned application of information theory to biochemical reaction
networks to show the limits on the ability to control fluctuations
in molecular abundances inside the cells as a function of the number
of signaling events~\cite{les10}.

This shows that a deeper knowledge of the thermodynamics and the
statistical physics of feedback controlled systems will bring a
deeper understanding of nanotechnology devices and biological
systems, and consequently it will increase our capacity of designing
and controlling them.

%%%%%%%%%%%%%%%%%%%%%%%%%%%%%%%%%%%%%%%%%%%%%%%
\section{Conclusions} \label{sec:conclusions}
%%%%%%%%%%%%%%%%%%%%%%%%%%%%%%%%%%%%%%%%%%%%%%%

Feedback or closed-loop control is present whenever there is a
controller or external agent that gathers information on the state
of the system and uses this information to decide the action it will
perform on the system. Feedback control allows to increase the
performance, and it is present in many systems with interest for
physicists, engineers and biologists~\cite{bec05}.

We have briefly reviewed the main ideas of the historical results by
Landauer~\cite{lan61}, Bennett~\cite{ben82}, and
Zurek~\cite{zur89,zur89b}, as well as the more recent results on the
limits to the control of a general system~\cite{tou00,tou04}, the
limits to the performance of feedback controlled
ratchets~\cite{cao07,fei07}, the thermodynamics of feedback
controlled systems~\cite{cao09}, and the generalization of
statistical physics identities and relations to feedback controlled
systems~\cite{sag12,sag12-1,sag12-2,sag12-3,abr12,hor10,pon10,lah12}.
This brief review has helped us to give the context for the open
questions that we have presented in Section~\ref{sec:open}. These
questions concern the quantification of the limits to the
performance increase that can be obtained thanks to a feedback that
uses a certain amount of information and the establishment of the
thermodynamics of feedback controlled systems. Answering these
questions is an important task for the theoretical understanding of
feedback controlled systems, which has important technological
applications. Feedback flashing
ratchets~\cite{cao04,cao04-1,cao04-2,cao04-3,cao04-4,cao04-5,cao04-6,cra08,cra08-1,lop08}
and Markovian particle pumps~\cite{cao09,toy10} are examples of
paradigmatic motors whose deeper understanding will benefit the
comprehension of the biological and technological devices they
model, and boost new applications. Control engineers have also
addressed similar topics from the point of view of dynamical
systems~\cite{mit05,san07} using models and techniques of optimal
control theory~\cite{bec05,ste94} as the Linear-Quadratic-Gaussian
control or the Kalman filter. The development of the links between
physics, information theory and control theory will allow to
quantify and to establish constraints to the increase of performance
reachable with a given amount of information in general feedback
controlled systems.

\acknowledgments

We acknowledge financial support from MCYT (Spain) through the
Research Projects FIS2006-05895, FIS2010-17440, from the ESF
Programme STOCHDYN, and from UCM and CM (Spain) through
CCG07-UCM/ESP-2925 and 920911.


\begin{thebibliography}{0}
\bibitem{bec05} Bechhoefer, J. Feedback for physicist: A tutorial essay on
   control. {\em Rev. Mod. Phys.} {\bf 2005}, {\em 77}, 783--836.
\bibitem{lef03} Leff, H.S.; Rex, A.F. \emph{Maxwell's Demon 2: Entropy,
    Classical and Quantum Information, Computing}; Institute of Physics:
   Bristol, UK, 2003.
\bibitem{szi29} Szilard, L. On the decrease of entropy in a thermodynamic
   state by the intervention of intelligent beings. {\em Z. Phys.} {\bf 1929},
   {\em 53}, 840--856.
\bibitem{lan61} Landauer, R. Irreversibility and heat generation in the
   computing process. {\em IBM J. Res. Dev.} {\bf 1961}, {\em 5}, 183--191.
\bibitem{ben82} Bennett, C.H. The thermodynamics of computation---A review.
   {\em Int. J. Theor. Phys.} {\bf 1982}, {\em 21}, 905--940.
\bibitem{zur89} Zurek, W.H. Algorithmic randomness and physical entropy. {\em
    Phys. Rev. A} \textbf{1989}, {\em 40}, 4731--4751.
\bibitem{zur89b} Zurek, W.H. Thermodynamic cost of computation, algorithmic
  complexity and the information metric. {\em Nature} \textbf{1989}, {\em 341},
  119--124.
\bibitem{ben73} Bennett, C.H. Logical reversibility of computation. {\em IBM
    J. Res. Dev.} {\bf 1973}, {\em 17}, 525--532.
\bibitem{llo89} Lloyd, S. Use of mutual information to decrease entropy:
  Implications for the second law of thermodynamics. {\em Phys. Rev. A} {\bf
    1989}, {\em 39}, 5378--5386.
\bibitem{tou00} Touchette, H.; Lloyd, S. Information-theoretic limits of
  control. {\em Phys. Rev. Lett.} {\bf 2000}, {\em 84}, 1156--1159.
\bibitem{tou04} Touchette, H.; Lloyd, S. Information-theoretic approach to the
  study of control systems. {\em Physica~A} {\bf 2004}, {\em 331}, 140--172.
\bibitem{cov91} Cover, T.M.; Thomas, J.A. \emph{Elements of Information
    Theory}; John Wiley: New York, NY, USA, 1991.
\bibitem{cao09} Cao, F.J.; Feito, M. Thermodynamics of feedback controlled
  systems. {\em Phys. Rev. E} {\bf 2009}, {\em 79}, 041118.
\bibitem{les10} Lestas, I.; Vinnicombe, G.; Paulsson, J. Fundamental limits on
  the suppression of molecular fluctuations. {\em Nature} {\bf 2010}, {\em 467},
  174--178.
\bibitem{cao04} Cao, F.J.; Dinis, L.; Parrondo, J.M.R. Feedback control in a
   collective flashing ratchet. {\em Phys.~Rev. Lett.} {\bf 2004}, {\em 93},
   040603.
\bibitem{cao04-1}
  Dinis, L.; Parrondo, J.M.R; Cao, F.J. Closed-loop control strategy with
   improved current for a flashing ratchet. {\em Europhys. Lett.} {\bf 2005},
   {\em 71}, 536--541.
\bibitem{cao04-2}  Feito, M.; Cao, F.J. Threshold feedback control for a collective flashing
   ratchet: Threshold dependence. {\em Phys. Rev. E} {\bf 2006}, {\em 74},
   041109.
\bibitem{cao04-3}  Feito, M.; Cao, F.J. Time-delayed feedback control of a flashing ratchet.
   {\em Phys. Rev. E} {\bf 2007}, {\em 76}, 061113.
\bibitem{cao04-4}  Feito, M.; Cao, F.J. Transport reversal in a delayed feedback ratchet.
   {\em Physica A} {\bf 2008}, {\em 387}, 4553--4559.
\bibitem{cao04-5}  Feito, M.; Cao, F.J. Optimal operation of feedback flashing ratchets.
   {\em J. Stat. Mech. Theor. Exp.} {\bf 2009}, P01031, doi:10.1088/1742-5468/2009/01/P01031.
\bibitem{cao04-6}  Feito, M. {Feedback Brownian ratchets and information}. Ph.D. Thesis,
   Editorial Universidad Complutense de Madrid: Madrid, Spain, 2009. Available online:
   http://eprints.ucm.es/10680/ 1/T31799.pdf (accessed on 28 March 2012).
\bibitem{cra08} Craig, E.M.; Kuwada, N.J.; Lopez, B.J.; Linke, H.
   Feedback control in flashing ratchets. {\em Ann.~Phys.} {\bf 2008}, {\em 17},
   115--129.
\bibitem{cra08-1}
  Craig, E.M.; Long, B.R.; Parrondo, J.M.R.; Linke, H. Effect of time delay on
   feedback control of a flashing ratchet.  {\em Europhys. Lett.} {\bf 2008}, {\em 81},
   10002.
\bibitem{feedback} Gao, T.; Chen, J. The current transport characteristics of
   a delayed feedback ratchet in a double-well potential. {\em J. Phys. A:
   Math. Theor.}  {\bf 2009}, {\em 42}, 065002.
\bibitem{feedback-1}  Gao, T.-F.; Liu, F.-S.; Chen, J.-C. Feedback control in a coupled Brownian
   ratchet. {\em Chin. Phys. B} {\bf 2012}, {\em 21}, 020502.
\bibitem{inertial} Son, W.-S.; Ryu, J.-W.; Hwang, D.-U.; Lee, S.-Y.;
   Park, Y.-J.; Kim, C.-M. Transport control in a deterministic ratchet
   system. {\em Phys. Rev. E} {\bf 2008}, {\em 77}, 066213.
\bibitem{inertial-1}  Hennig, D.; Schimansky-Geier, L.; H\"anggi, P. Directed transport of an
   inertial particle in a washboard potential induced by delayed feedback. {\em
    Phys. Rev. E} {\bf 2009}, {\em 79}, 041117.
\bibitem{inertial-2}  Hennig, D. Current control in a tilted washboard potential via time-delayed
   feedback. {\em Phys. Rev.~E} {\bf 2009}, {\em 79}, 041114.
\bibitem{inertial-3}  Zhang, X.-M.; Ai, B.-Q. Transport of overdamped Brownian particles driven
   by ac forces and time-delayed feedback. {\em J. Phys. A: Math. Theor.} {\bf
   2010}, {\em 43}, 495004.
\bibitem{inertial-4}  Du, L.-C.; Mei, D.-C. Time delay control of absolute negative mobility and
   multiple current reversals in an inertial Brownian motor. {\em J. Stat. Mech.
   Theor. Exp.} {\bf 2011}, \emph{2011}, P11016.
\bibitem{lop08} Lopez, B.J.; Kuwada, N.J.; Craig, E.M.; Long, B.R.; Linke,
   H. Realization of a Feedback Controlled Flashing
   Ratchet. {\em Phys. Rev. Lett.} {\bf 2008}, {\em 101}, 220601.
\bibitem{ast98} Astumian, R.D.; Der\'enyi, I. Fluctuation driven transport and
   models of molecular motors and pumps. {\em Eur. Biophys. J.} {\bf 1998},
   {\em 27}, 474--489.
\bibitem{zho96} Zhou, H.-X.; Chen, Y.-D. Chemically Driven Motility of
   Brownian Particles. {\em Phys. Rev. Lett.} {\bf 1996}, {\em 77}, 194--197.
\bibitem{bie07} Bier, M. The Stepping Motor Protein as a Feedback Control
   Ratchet. {\em Biosystems} {\bf 2007}, {\em 88}, 301--307.
\bibitem{ser07} Serreli, V.; Lee, C.-F.; Ray, E.R.; Leigh, D. A molecular
  information ratchet. {\em Nature} {\bf 2007}, {\em 445}, 523--527.
\bibitem{fey63} Feynman, R.P.; Leighton, R.B.; Sands, M. \emph{The Feynman
    Lectures on Physics I}; Addison-Wesley: Reading, MA, USA, 1963; Chapter 46.
\bibitem{bro05} Van den Broeck, C.; Meurs, P.;  Kawai, R. From Maxwell demon to Brownian motor. \emph{New J. Phys.} {\bf
    2005}, \emph{7}, 10.
\bibitem{cao07} Cao, F.J.; Feito, M.; Touchette, H. Information and flux in a
   feedback controlled Brownian ratchet. {\em Physica A} {\bf 2007}, {\em
     388}, 113--119.
\bibitem{fei07} Feito, M.; Cao, F.J. Information and maximum power in a
  feedback controlled Brownian ratchet. {\em Eur. Phys. J. B} {\bf 2007}, {\em
   59}, 63--68.
\bibitem{all08} Allahverdyan, A.E.; Saakian, D.B. Thermodynamics of adiabatic
   feedback control. {\em Europhys.~Lett.} {\bf 2008}, {\em 81}, 30003.
\bibitem{toy10} Toyabe, S.; Sagawa, T.; Ueda, M.; Muneyeki, E.; Sano, M.
  Experimental demonstration of information-to-energy conversion and
  validation of the generalized Jarzynski equality. {\em Nature~Phys.} {\bf
    2010}, {\em 6}, 988--992.
\bibitem{abr11} Abreu, D.; Seifert, U. Extracting work from a single heat bath
   through feedback. {\em Europhys. Lett.} {\bf 2011}, {\em 94}, 10001.
\bibitem{abr11-1}  Horowitz, J.M.; Parrondo, J.M.R. Thermodynamic reversibility in feedback
   processes. {\em Europhys.~Lett.} {\bf 2011}, {\em 95}, 10005.
\bibitem{abr11-2}  Vaikuntanathan, S.; Jarzynski, C. Modeling Maxwell’s demon with a
  microcanonical Szilard engine. {\em Phys. Rev. E} {\bf 2011}, {\em 83},
  061120.
\bibitem{abr12} Abreu, D.; Seifert, U. Thermodynamics of genuine nonequilibrium
  states under feedback control. {\em Phys. Rev. Lett.} {\bf 2012}, {\em 108},
  030601.
\bibitem{sag12} Sagawa, T.; Ueda, M. Nonequilibrium thermodynamics of feedback
   control. {\em Phys. Rev. E} {\bf 2012}, {\em 85}, 021104.
\bibitem{sag12-1}  Sagawa, T.; Ueda, M. Generalized Jarzynski equality under nonequilibrium
   feedback control. {\em Phys. Rev. Lett.} {\bf 2010}, {\em 104}, 090602.
\bibitem{sag12-2}  Sagawa, T. Hamiltonian derivations of the generalized Jarzynski equalities
   under feedback Ccontrol. {\em J. Phys. Conf. Ser.} {\bf 2011}, {\em 297},
   012015.
\bibitem{sag12-3}  Sagawa, T. Thermodynamics of information processing in small systems. {\em
    Prog. Theor. Phys.} {\bf 2012}, {\em 127}, 1--56.
\bibitem{jar97} Jarzynski, C. Nonequilibrium equality for free energy
   differences. {\em Phys. Rev. Lett.} {\bf 1997}, {\em 78}, 2690--2693.
\bibitem{hor10} Horowitz, J.M.; Vaikuntanathan, S. Nonequilibrium detailed
   fluctuation theorem for repeated discrete feedback. {\em Phys. Rev. E} {\bf
    2010}, {\em 82}, 061120.
\bibitem{pon10} Ponmurugan, M. Generalized detailed fluctuation theorem under
   nonequilibrium feedback control. {\em Phys. Rev. E} {\bf 2010}, {\em 82},
   031129.
\bibitem{lah12} Lahiri, S.; Rana, S.; Jayannavar, A.M. Fluctuation theorems in
   the presence of information gain and feedback. {\em J. Phys. A:
   Math. Theor.} {\bf 2012}, {\em 45}, 065002.
\bibitem{mit05} Mitter, S.K.; Newton, N.J. Information and entropy flow in the
   Kalman–Bucy filter. {\em J. Stat. Phys.} {\bf 2005}, {\em 118}, 145--176.
\bibitem{san07} Sandberg, H.; Delvenne, J.-C.; Doyle,
  J.C. Linear-Quadratic-Gaussian heat engines.\ {In Proceedings of the 46th IEEE Conference on
  Dec. and Control}, New Orleans, LA, USA, 12--14 December {2007}; pp. 3102--3107.
\bibitem{ste94} Stengel, R.F. \emph{Optimal Control and Estimation}; Dover
  Publications: New York, NY, USA, 1994.
\end{thebibliography}
\end{document}